\documentstyle[epsfig]{mn}
\newif\ifAMStwofonts
\newcommand{\be}{\begin{equation}}
\newcommand{\ee}{\end{equation}}
\newcommand{\bea}{\begin{eqnarray}}
\newcommand{\eea}{\end{eqnarray}}
\newcommand{\nn}{\nonumber}

\newcommand{\etal}{ {\it et al.} }

\newcommand{\W}{{\bf W}}
\newcommand{\Wr}{W_r}
\newcommand{\Wth}{W_\theta}
\newcommand{\Wphi}{W_\phi}
\newcommand{\br}{b_r}
\newcommand{\Br}{B_r}
\newcommand{\bth}{b_\theta}
\newcommand{\Bth}{b_\theta}
\newcommand{\bbphi}{b_\phi}

\newcommand{\dth}{d_\theta}
\newcommand{\sinth}{\sin\theta}
\newcommand{\p}{p}
\newcommand{\bphimax}{b_{\phi,max}}
\newcommand{\brmax}{b_{r,max}}

\newcommand{\ma}{m_{A,\theta}}

\newcommand{\Alfven}{Alfv\'en }
\newcommand{\Alfvenic}{Alfv\'enic }
\newcommand{\Brout}{\Br(0)}
\newcommand{\Brin}{\Br(\pi/2)}
\newcommand{\Wrout}{\Wr(0)}
\newcommand{\Wrin}{\Wr(\pi/2)}
\newcommand{\bphiturn}{\br(\theta_0)}
\newcommand{\bpolturn}{b_{pol}(\theta_0)}

\newcommand{\psibar}{\bar{\psi}}
\newcommand{\mubar}{\bar{\mu}}
\newcommand{\bbar}{\bar{\bf b}}
\newcommand{\brbar}{\bar{b}_r}
\newcommand{\bthbar}{\bar{b}_\theta}
\newcommand{\bphibar}{\bar{b}_\phi}
\newcommand{\Wbar}{\bar{\bf W}}
\newcommand{\Wrbar}{\bar{W}_r} 
\newcommand{\Wthbar}{\bar{W}_\theta}
\newcommand{\Wphibar}{\bar{W}_\phi}
\newcommand{\pbar}{\bar{p}}
\newcommand{\Erbar}{\bar{E}_r}
\newcommand{\Ethbar}{\bar{E}_\theta}
\newcommand{\Ephibar}{\bar{E}_\phi}

\title{Protostellar Evolution during Time Dependent, Anisotropic Collapse }

\author[ Mahmoud Aburihan, J.D. Fiege, R.N. Henriksen \& T. Lery]
{Mahmoud Aburihan$^1$ \thanks{This paper developed from an insightful MSc. thesis by
Mahmoud Aburihan whose accidental death on Dec. 20, 1999
is greatly regretted by his colleagues and many friends.
His co-authors would like to dedicate this work to his memory
and to his family.
},  
Jason D. Fiege$^2$\thanks{fiege@cita.utoronto.ca},
Richard N. Henriksen$^1$\thanks{henriksn@astro.queensu.ca}, and
Thibaut Lery$^3$\thanks{lery@cp.dias.ie}\\
$^1$Queen's University, Kingston, Ontario,Canada \\
$^2$Canadian Institute for Theoretical Astrophysics,
McLennan Labs, University of Toronto 60 St. George Street, Toronto, Ontario, M5S 3H8 \\
$^3$ DIAS, School of Cosmic Physics, 5 Merrion Square, Dublin 2 Ireland
}

\date{\today}
\pagerange{\pageref{firstpage}--\pageref{lastpage}}
\pubyear{2000}

\begin{document}

\maketitle

\label{firstpage}

\begin{abstract}
The formation and collapse of a protostar involves the simultaneous infall and outflow of material 
in the presence of magnetic fields, self-gravity, and rotation.  We use self-similar techniques to 
self-consistently model the anisotropic collapse and outflow by a set of angle-separated self-similar equations. 
The outflow is quite strong in our model, with the velocity increasing in proportion to radius, and material
formally escaping to infinity in the finite time required for the central singularity
to develop.

Analytically tractable collapse models have been limited mainly to spherically symmetric
collapse, with neither magnetic field nor rotation.  Other analyses usually employ extensive 
numerical simulations, or either perturbative or quasistatic techniques.  Our model is unique as 
an exact solution to the non-stationary equations of self-gravitating MHD, which features
co-existing regions of infall and outflow.

The velocity and magnetic topology of our model is quadrupolar, although dipolar solutions
may also exist.  We provide a qualitative model for the origin and subsequent evolution of 
such a state.  However, a central singularity forms at late times, and we expect the late time behaviour 
to be dominated by the singularity rather than to depend on the details of its initial state.  Our solution may, 
therefore, have the character of an attractor among a much more general class of self-similarity.

%------------------------ 
%       Abstract 
%------------------------ 
     
\end{abstract} 

\section{Introduction}
\label{sec:intro}

\renewcommand{\textfraction}{0}
\renewcommand{\topfraction}{1} 
\renewcommand{\bottomfraction}{1}

It is clear that outflows often co-exist with infall as protostars form within 
the collapsing cores of molecular clouds (Bertout 1989; Andr\'e \etal 1993).  Infall and outflow both appear to be 
present for much of the protostellar ``main sequence,'' from rapidly accreting embedded 
Class 0 objects to fully formed T Tauri stars.  This suggests that the dynamics leading to the 
formation of a protostar are more complex than simple radial infall, and are dominated 
by strongly anisotropic motions.  We present a new model for the anisotropic collapse of a molecular
cloud core, which self-consistently treats the effects due to self-gravity, magnetic fields,
and rotation, as the central protostellar core grows and a bipolar outflow develops.

Self-gravitating models of protostellar collapse have usually
been limited to the classical solutions with spherical
symmetry (Larson, 1969; Penston, 1969; Shu, 1977), including the 
elaborations and clarifications in related works (Hunter, 1977;
Whitworth and Summers, 1985; Henriksen, Andr\'e and Bontemps, 1997, hereafter HAB1997).  
Galli and Shu (1993, hereafter GS) presented a very
interesting calculation, which included the effects of anisotropy as a perturbation about the
classical Shu (1977) inside-out collapse solution.  Subsequently, Li and Shu (1996) presented a 
quasi-static calculation.  An approximate analytic self-similar treatment based on a dynamic
termination of the ambipolar diffusion models has also been given recently and
developed to the point of comparison with observations (e.g. Basu, 1997).  However
the bipolar outflow was not integral to any of these papers, as it is in the case of
the present work.

We have previously studied steady-state solutions for simultaneous infall and outflow 
late in the evolutionary sequence, after the dominant central mass had already formed
(Henriksen and Valls-Gabaud, 1994; Fiege \& Henriksen 1996, hereafter FH1; Lery, Henriksen, \& Fiege 1999, hereafter LHF).
Thus, our previous models apply only to very late times in the formation of 
a protostar.  The philosophy of these articles was that bipolarity could be studied
using scale free solutions near the central singularity without
worrying too much about the initial state of the flow.  This is similar
in spirit to the development of the Larson-Penston self-similar
solution from non-self-similar initial conditions.  

The model presented in this paper takes quite a different approach by
treating the time-dependent problem of accretion and simultaneous outflow in
a dynamically collapsing and self-gravitating core.
Thus, we study an early stage of stellar formation when
the star has not yet formed, and most of the gas still resides in the
surroundings. A limit to the self-gravitating regime is certainly set
when the mass of the central object dominates that of the
surroundings. Thus our present solution is a natural complement to our
earlier studies.

The present model is best described as an inner ``settling'' solution, which follows the assembly of the protostar in detail.
It is limited to smaller spatial scales than our previous work, but
the flow structure that we predict would presumably be embedded within a larger collapsing region.
This larger region might include nearly steady-state outflows and an accretion region of the self-similar type that 
we have previously discussed in FH1 and LHF (See Section \ref{sec:discussion}).  

Numerical simulations have also been used to study non-isotropic self-gravitating collapse.
For example, Tomisaka (1998) used a multi-grid MHD code to study the gravitational collapse of
a segment of  a magnetized and slowly rotating filament.  His simulation resulted in a 
rotating pseudo-disc, which produced an outflow as the central object grew.  Most interesting, from
our point of view, is that the late stages of the calculation were dominated by a quadrupolar velocity field.
Such quadrupolarity can develop in super-\Alfvenic flows when infalling material near the
midplane is deflected up the axis due to a combination of
pressure gradients, magnetic fields, and the centrifugal barrier.
Our present calculation shows that this mechanism
can operate on smaller scales between the growing boundary of 
the hydrostatic core and the pressure dominated region external to the region of self-similarity.
The development of quadrupolar structure and the connection of our model to the 
exterior region is discussed in Section \ref{sec:discussion}.

Our central assumption is that this early stage of star formation is dynamic rather
than quasi-static.  This is suggested by various lines of evidence
(Basu, 1997; Foster and Chevalier, 1993; HAB, 1997), 
although this assumption cannot be regarded as 
certain (Basu and Mouschovias 1995a; Basu and Mouschovias 1995b;
Galli and Shu, 1993; Li and Shu,1997). In any case, the central regions must
ultimately become hydrostatic to allow for the growth of
the naissant protostar.  Thus, we expect and do indeed find ``settling'' solutions,
in which the radial velocity goes to zero
near the centre, where the infalling material actually forms the stellar core.

In the following section, we derive our basic self-similar equations from self-gravitating MHD,
under the assumption of self-similar flow.  We show, in Section \ref{sec:Analysis}, that our equations admit an 
exact and completely analytic class of solutions.  We use these solutions to derive several interesting 
analytic results, which illustrate and constrain the properties of the model. 
Our most important result is that substantial outflow velocities can be obtained without resorting to 
heating the material, as in FHI and LHF.  
The axial outflow velocity is never more than twice the equatorial inflow velocity
on a sphere of some given radius, at some instant of time.  
However, we find that the outflow velocity increases linearly with distance,
so that substantial velocities are obtained far from the origin.  
By following the motion of individual fluid elements, we find that all such 
fluid elements escape to radial infinity in the finite time required for the central singularity to develop.
This escaping gas would presumably interact with the external medium in a complicated manner, which we discuss in Section \ref{sec:discussion}.  Our Discussion section also presents a simple, qualitative model to 
provide one possible way in which the quadrupolar structure might arise.  We also note that models with dipolar
geometry are also possible, although we do not find them explicitly in the present calculation.

Finally, we note that a preliminary account of this work was presented at the Cracow meeting on 
"Plasma Turbulence and Energetic Particles in Astrophysics"(1999, M. Ostrowski and R. Schlickeiser, eds.).

\section{Self-Similar Accretion and Outflow}
\label{sec:origin}

We assume that the protostellar development begins with the primarily
radial collapse of a cold molecular core.  In reality, cores are not really expected
to be spherical (Tomisaka \etal 1988, Myers \etal 1991, Ryden 1996, Fiege \& Pudritz 2000a), but
all that is really required is that there be a substantial collapse in a plane perpendicular to
the initial magnetic and rotation axes (assumed to be parallel for simplicity).  The core may have been in
equilibrium initially as a singular isothermal sphere (SIS) as in GS, or it may have had a more
complicated internal structure as suggested by HAB.  The initial conditions
provide a set of characteristic scales, which suggest the set of 
conserved quantities used to define the class of self-similar symmetry.
For example, the obvious constants for an unbounded SIS are the sound speed $c_s$ and
Newton's constant $G$.  GS used these constants to define a ``class'' of
self-similarity obeyed during the collapse.  On the other hand, a constant external pressure
bounding a truncated self-gravitating sphere, together with $G$,
would define a different class of self-similarity.  Yet another similarity class would
be appropriate if there were a characteristic time due to rotation.

Carter and Henriksen (1991) developed a mathematical formalism
for determining the most general class of self-similarity possible, which naturally includes all possible
initial states.  This method has been used successfully in studying the evolution of
collisionless n-body systems (see e.g. Henriksen, 1997), but is equally well-suited for 
studying the dynamical collapse of a magnetized, self-gravitating core.  This technique can
be used to demonstrate the existence of the separable (in radius, poloidal angle, and time)
``settling'' solution presented here, as a special case within a more general class of
self-similarity.  It is possible that these separable solution might represent an ``attractor'' 
within this larger class of self-similar models (Aburihan, 1999), but we do not prove this here.

It is important to address the question of how the quadrupolar magnetic and flow structure
assumed by our model might originate.  Such questions are really beyond the scope
of our self-similar treatment, since self-similar solutions are often {\em intermediate}, in the
sense that they are disconnected from their boundary conditions in space or time (see discussion in FH1).
Thus, our ideas regarding the origin of our self-similar model are somewhat speculative.
Nevertheless, we suggest a reasonable scenario in Section \ref{sec:discussion},
by which a rotating and magnetized cloud could evolve a quadrupolar flow and magnetic field
structure during collapse.  This would arise as a consequence of poloidal pressure gradients,
the centrifugal barrier encountered by the collapsing cloud, and magnetic reconnection effects
which change the topology of the field in regions where localized field reversals occur.
We discuss the details of this scenario in Section \ref{sec:discussion}, and
turn now to a derivation of the basic equations.

\subsection{Equations}
\label{sec:equations}
The basic non-dimensional quantities from which we construct our self-similar model are given by the 
poloidal angle $\theta$ and the variable
\be
X\equiv -\frac{r}{c_s~ t},
\label{eq:Xdef}
\ee
where $r$ is spherical radius, $t$ is time, and $c_s$ is a {\em fiducial} sound
speed.  The local sound speed need not be constant; the constant $c_s$ in equation
\ref{eq:Xdef} is only meant to be typical of the initial conditions.
The minus sign is included to make $X$ positive definite, since our model
starts with $t$ large and negative, evolving toward a singularity at $t=0$.  Note that
our $X$ is essentially the same as the self-similar variable used by Shu (1977) and
more recently by GS in the context of collapsing isothermal spheres,
although their model is strictly isothermal.  We provide a self-similar framework in this 
section which would, in principle, allow one to follow the collapse of a SIS.  

Our treatment admits considerable freedom in choosing the equation of state (EOS).  
One particularly simple non-isothermal choice, which we discuss
later in this section and use extensively throughout this paper, makes the partial differential
equations separable, which allows us to find a singular and
completely analytic solution to our self-similar equations.  Unfortunately, this particular
solution does not match on to the SIS at early times.  A more complete treatment of the
collapse problem would solve (with greater effort) the general self-similar PDEs directly, 
beginning with realistic initial conditions, such as a SIS threaded by a magnetic field, rather than seeking 
out special separable forms.  However, our singular solution might 
represent an ``attractor'' among the more general class of self-similarity, which would make
it a valid end state for a wide variety of initial conditions.  This possibility is 
further explored in Section \ref{sec:discussion}.
 
The actual temperature and sound speed in our model vary as functions of $r$, $\theta$, and $t$
in a way that is uniquely determined by the self-similarity.  Generally speaking,
the gas heats up during collapse in our model, although with significant temperature structure in $\theta$.
Isothermality in cores is maintained primarily by the efficient cooling
provided by molecular lines.  An anisotropically collapsing core would likely evolve toward 
a more complex temperature structure once the dynamical timescale $r/c_s$ becomes shorter 
than the cooling time.  Thus, we expect the self-similarity to develop from the inside out,
with the central regions of the collapsing core evolving more rapidly toward a self-similar flow pattern,
while the outer layers remain nearly isothermal until later in the collapse.

The most general class of self-similarity based on $X$ and $\theta$ is expressed
by writing all physical quantities as functions of these variables:
\bea 
\rho &=& \frac{c_s^2}{4\pi G r^2} \mubar(X,\theta) \label{eq:rhoform}\\
{\bf v} &=& c_s \Wbar(X,\theta) \label{eq:vform}\\
{\bf B} &=& \sqrt{\frac{c_s^4}{G}} \frac{\bbar(X,\theta)}{r} \label{bform}\\
\Phi &=& c_s^2 \psibar(X,\theta) \label{eq:psiform}\\
P &=& \frac{c_s^4}{4\pi G} \left(\frac{p_0(t)}{r_0^2}+\frac{\pbar(X,\theta)}{r^2} \right). \label{eq:pform}
\eea
The notation for the dimensional magnetohydrodynamic (MHD) quantities on the left land side of these equations is standard, and 
the quantities $\mubar$, $\Wbar$, $\bbar$, $\psibar$, and $\pbar$ on the right hand side are dimensionless
forms of the density, velocity, magnetic field, gravitational potential, and ``dynamical'' pressure respectively.
Note that we separate the pressure into a ``dynamical'' pressure term $\pbar$, and a background pressure term $p_0(t)$.
The term involving $p_0(t)$ has no dynamical consequences whatsoever, since it does not contribute to the pressure gradient.
It is necessary because realistic solutions are obtained only when the ``dynamical'' part of the pressure $\pbar<0$,
as we shall further discuss in Section \ref{sec:method}.
Finally, note that the overbars distinguish these quantities from the separated forms of these variables, presented later in this section.

We now use the preceding ansatz of self-similarity in each of the equations of self-gravitating MHD to 
obtain our general set of self-similar equations.  Poisson's equation, the continuity equation, and the condition that
there are no magnetic monopoles are written respectively as follows:\\
{\bf Poisson's Equation}
\be
X^2\partial_X^2\psibar+2X\partial_X\psibar+\frac{1}{\sin\theta}\partial_\theta(\sin\theta\partial_\theta\psibar)=\mubar~;
\label{poiss}
\ee
{\bf Continuity}
\be
\label{mass}
(\Wrbar + X)X\partial_X \mubar+\mubar X\partial_X
\Wrbar+\frac{1}{\sin\theta}\partial_\theta(\mubar \Wthbar \sin\theta)=0;
\ee 
{\bf No Magnetic Poles}
\be
\label{nopole}
\partial_X(X \brbar)+\frac{1}{\sin\theta}\partial_\theta(\sin\theta \bthbar)=0.
\ee
The 3 components of the MHD induction equation are given by the following:\\
{\bf Induction Equation - $\hat{r}$}
\be
\label{rflux}
- X^2\partial_X \brbar=\frac{1}{\sin\theta}\partial_\theta(\sin\theta \Ephibar)~;
\ee
{\bf Induction Equation - $\hat{\theta}$}
\be
\label{tflux}
- X\partial_X \bthbar +\partial_X \Ephibar=0~;
\ee 
{\bf Induction Equation - $\hat{\phi}$}
\be
\label{phflux}
- X\partial_X \bphibar=\partial_X \Ethbar-\frac{1}{X}\partial_\theta \Erbar, 
\ee
where $\Erbar$, $\Ethbar$, and $\Ephibar$ are related to the electric field and given by
\bea
\Ephibar &\equiv& -(\Wrbar \bthbar-\Wthbar \brbar),\\
\Ethbar &\equiv& -(\Wphibar \brbar-\Wrbar \bphibar),\\
\Erbar  &\equiv& -(\Wthbar \bphibar-\Wphibar \bthbar).
\eea
Finally, the 3 components of the momentum equation are given as follows:\\
{\bf  Momentum - $\hat{r}$}
\bea
&& X(\Wrbar + X)\partial_X \Wrbar + \Wthbar\partial_\theta
\Wrbar-(\Wthbar^2+\Wphibar^2) \nonumber\\
&& =\frac{2\pbar}{\mubar}-\frac{1}{\mubar}X\partial_X \pbar
-X\partial_X \psibar+\frac{1}{\mubar} \bthbar \partial_\theta
\brbar \nn\\
&& -\frac{1}{2\mubar}X\partial_X(\bthbar^2+\bphibar^2)~;
\label{rmom}
\eea
{\bf Momentum - $\hat{\theta}$}
\bea
&& X(\Wrbar + X)\partial_X \Wthbar + \Wthbar\partial_\theta \Wthbar
+\Wrbar \Wthbar - \Wphibar^2 \cot\theta \nonumber\\
&& = -\frac{1}{\mubar}\partial_\theta \pbar - \partial_\theta\psibar
+\frac{1}{\mubar}\brbar X\partial_X \bthbar
-\frac{1}{\mubar} \bphibar^2 \cot\theta \nn\\
&& -\frac{1}{2\mubar}
\partial_\theta(\brbar^2+\bphibar^2);
\label{tmom} 
\eea
{\bf Momentum - $\hat{\phi}$}
\bea
&& X(\Wrbar + X)\partial_X \Wphibar + \frac{1}{\sin\theta} \Wthbar \partial_\theta
(\sin\theta \Wphibar) + \Wphibar \Wrbar \nn\\
&& =\frac{1}{\mubar}\brbar X\partial_X \bphibar +
\frac{1}{\mubar\sin\theta}\bthbar\partial_\theta(\sin\theta \bphibar).
\label{phmom}
\eea

Note that the self-similar forms presented in equations \ref{eq:rhoform} to \ref{eq:pform} would represent a 
SIS if $\mubar=2$ and $\Wbar=0$.  
A complete treatment of the self-similar problem would involve solving the ($X$,$\theta$) partial differential
equations presented above, starting from a slightly perturbed SIS at some initial time $t_0$ ($t_0=-\infty$ for an infinitestimal
perturbation).  In principle, this approach would allow us to follow the entire
anisotropic development of the SIS all the way to the development of the singularity at $t=0$.  This could be
done under various assumptions regarding the initial conditions for the rotation and magnetic field, resulting in a very
complete model of protostellar collapse.  This ambitious project will be left for future work.
Instead, we take a more modest approach in this
paper, by solving the restricted problem where all of the variables $\bar{f_i}$ written with overbars in 
equations \ref{eq:rhoform} to \ref{eq:pform} can be written in the separable form
\be
\bar{f_i}(X,\theta)=k_iX^{\alpha_i} f_i(\theta),
\label{eq:sep}
\ee
where the $k_i$ are appropriate normalizing constants. In the case of
$\bar{\mu}$ and $\bar{p}$ these constants are each $4\pi$,. They
are equal to $+1$ elsewhere.   
The factor that depends on $X$ is eliminated by carefully balancing powers of $\alpha_i$, 
resulting in the following system of ordinary differential equations
in $\theta$, written here in the same order as in equations \ref{poiss} to \ref{phflux} and \ref{rmom} to \ref{phmom}. 
\be
6\Psi + \frac{1}{\sinth} \dth (\sinth \dth\Psi) = 4\pi\mu,
\label{eq:Psi} 
\ee
\be
3 \Wr+2 +\frac{1}{\sinth} \dth(\sinth \Wth) + \Wth\dth\ln\mu= 0,
\label{eq:continuity}
\ee
\be
3\br +\frac{1}{\sinth} \dth(\sinth \bth) = 0,
\label{eq:divb}
\ee
\be
-2\br+\frac{1}{\sinth} \dth\left[\sinth(\Wr\bth-\Wth\br)\right] = 0,
\label{eq:ind1}
\ee
\be
2\bth-3(\Wth\br-\Wr\bth) = 0,
\label{eq:ind2} 
\ee
\be
-2\bbphi+3(\Wphi\br-\Wr\bbphi)+\dth(\Wphi\bth-\Wth\bbphi) = 0,
\label{eq:ind3}
\ee
\bea
&& \Wr+\Wr^2+\Wth\dth\Wr-(\Wth^2+\Wphi^2) \nn\\ 
&=& -2\frac{p}{\mu}-2\Psi+\frac{1}{4\pi\mu}\left[(\bth\dth\br-2(\bth^2+\bbphi^2)\right], 
\label{eq:mom1}
\eea
\bea
&& \Wth +\Wth\dth\Wth +2\Wr\Wth-\Wphi^2\cot\theta = -\frac{\dth p}{\mu}-\dth\Psi \nn\\
&+& \frac{1}{4\pi\mu}\left[2\br\bth-\frac{1}{2}\dth(\br^2+\bbphi^2)-\bbphi^2\cot\theta\right] 
\label{eq:mom2}
\eea
\bea
&& \Wphi+2\Wr\Wphi +\frac{\Wth}{\sinth}\dth(\sinth \Wphi) \nn\\
&=& \frac{1}{4\pi\mu}\left[2\br\bbphi+\frac{\bth}{\sinth}\dth(\sinth\bbphi)\right]. 
\label{eq:mom3}
\eea
Equations \ref{eq:Psi} to \ref{eq:mom3} are the final equations, whose solutions we study 
for the remainder of this paper. 

The physical variables are given explicity by the following forms, 
in which we have replaced $X$ by $r$ and $t$, using the definition of $X$ given
in equation \ref{eq:Xdef}:
\bea
{\bf v}&=& -\frac{r}{t}{\bf W}(\theta), \label{eq:vform2}\\
{\bf B}&=& \frac{r}{\sqrt{G}t^2}{\bf b}(\theta), \label{eq:bform2}\\
\rho &=& \frac{1}{G t^2}\mu(\theta), \label{eq:rhoform2}\\
p &=& \frac{1}{G t^4}\left[r_0^2 p_0^\ast +r^2 p(\theta) \right] \label{eq:pform2}\\
\Phi &=& \frac{r^2}{t^2}\Psi(\theta) \label{eq:Phiform2}.
\eea
Note that we have replaced the term involving $p_0(t)$ in equation \ref{eq:pform} with one involving only the 
constant $p_0^\ast$, which can be chosen {\it a posteriori} to keep the total pressure positive throughout the region
of interest. Formally we have set $p_o(t)=
\frac{4\pi}{c_s^4 t^4}r_o^4 p_o^\ast$.    

One should realize that our final set of equations can also be
obtained by using these physical forms directly 
in the equations of self-gravitating MHD, without ever writing down the ($X$, $\theta$) forms of the equations
(equations \ref{poiss} to \ref{phmom}).  However, our treatment has the advantage of providing the equations within the
context of a broader class of self-similarity, which may be useful for future analysis.
It is important to realize that $c_s$ has disappeared from these equations, as would any
dimensional constant used in equation \ref{eq:Xdef}. Thus there is a ``route'' to these
equations from quite general initial conditions, which is the main reason for our suspicion that 
it may be an attractor.

The nature of our model as an internal settling 
solution is clear from the self-similar forms in equations \ref{eq:vform2} to \ref{eq:Phiform2},
before we even solve the equations.  We observe that all velocity components go to zero
proportionally to $r$, which also implies that there is rigid rotation
on each cone defined by $\theta=const$.  Of course, the angular velocity is allowed to 
vary between cones, and material flows between them. 

A difficulty arises when one tries to solve the self-similar form of Poisson's equation given in
equation \ref{eq:Psi}.  Any solution to equation \ref{eq:Psi} must satisfy the boundary conditions
$\Psi'(0)=0$ and $\Psi'(\pi/2)=0$, so that there is no $\theta$ component of the gravitational 
acceleration at either boundary.
The solution to the homogeneous form of the equation ($\mu=0$) which satisfies these boundary conditions, 
is the following:
\be
\Psi_{hom}=c P_2(\cos\theta)=c \frac{1}{2} (3\cos^2\theta-1),
\ee
where $P_2(\cos\theta)$ is the second order Legendre polynomial and $c$ is an arbitrary constant.
Note that the homogeneous solution, with arbitrary $c$ may be added to any particular solution of equation \ref{eq:Psi},
which produces an additional term in the gravitational acceleration.  The solution to Poisson's equation is non-unique
for our system, since any such solution satisfies the boundary conditions at the polar and equatorial boundaries.
This non-uniqueness originates from the self-similarity.  Poisson's equation has a unique solution provided that 
appropriate boundary conditions can be specified {\em everywhere} on a surface enclosing the region of interest.
However, there is no unique way to specify the potential at radial
infinity for our distribution of matter, which in principle 
extends to infinity. In practice our solution must be halted at some
surface $R(\theta,t)$ whose form depends on the matching to an
external medium.  The best that we can do is to limit ourselves to the special case where $\mu(\theta)$ is constant and
the density distribution is spherically symmetric at all times.  In that case, $\Psi$ should be contant on physical
grounds, so that the isopotential surfaces are spherical as well.  Poisson's equation (\ref{eq:Psi}) is then trivially solved:
\be
\Psi=\frac{2}{3}\pi\mu.
\label{eq:Psi2}
\ee
We use this equation and the underlying assumptions of constant $\mu$ and $\Psi$ for the 
remainder of the analysis in this paper. 

Note that the restriction that $\mu=const$ implies a sort of ``poloidal incompressibility'' on our system, which takes the
place of a thermodynamic equation of state (EOS) in our system, in the same way that true incompressibility
replaces a thermodynamic EOS in incompressible fluid dynamics.  
It is generally not possible to impose any additional
EOS directly relating $p(\theta)$ to $\mu(\theta)$ or any other
variable, without imposing complete boundary conditions on Poisson's equation.

We commented in Section \ref{sec:origin} that our solution must only be valid 
near the centre of a collapsing core, where the dynamical timescale is much 
shorter than the cooling time of the molecular gas.  Realistically, our settling solution must be 
embedded within a nearly isothermal exterior region, which would remain nearly isothermal,
as a result of the efficient cooling provided by molecular lines and the longer dynamical timescale
outside of the collapsing region.  The boundary joining our model to such a region
would undoubtedly be complex in both shape and internal structure, with shocks arising near the
outflow axis.  A more complete model, which explicitly includes this external region would be very difficult to treat
analytically.  The equatorial region is expected to contain 
both acretion discs and magnetically neutral points as suggested
above, while there may be more violent activity (including shocks in
super Alfv\`enic flow) near the axes.  None of this can appear in the
simple asymptotic forms that we study here.  
The only effect of the external region in the present calculation is through the background pressure 
$p_0(t)$ in equation \ref{eq:pform2}, which has no effect on the dynamics whatsoever.

The self-similar forms given by equations \ref{eq:vform2} to \ref{eq:Phiform2} become
singular at $t=0$ over the entire domain of validity, and they cannot
be continued beyond this singularity.  This is
unlike the usual point singularity (i.e. a ``core'' of finite mass) that forms at $t=r=0$ in the
spherically symmetric collapse models, across which the solution may be
continued into the core accretion phase. Such a solution (but with
appropriate asymmetry) would be external to the solution presented here.  

The inner boundary of our solution is the growing hydrostatic
core and realistically, we expect the self-similar solution to vanish altogether
before $t=0$. This can happen if the outer boundary is shrinking in
time as the inner boundary grows. We can make this plausible near the
equatorial plane by observing that outside the transition region we
can expect a mass flux $\propto c_s^3/G$, as given by Shu (1977).  At the ``boundary'' $R$
of our inner region equations (\ref{eq:vform}) and (\ref{eq:rhoform}) show
that the mass flux scales as $R^3/t^3$ and thus by equating the two
expressions we obtain $R\propto c_s t$. Consequently the outer
boundary of our solution may be expected to shrink onto the hydrostatic
core before $t\rightarrow 0$, thus removing the mathematical
singularity from the domain over which our solution is valid.         
  
We turn in the next section to an analysis of these
equations.

\section{Integrals}
\label{sec:integrals}
There is considerable redundancy in our equations.  We note that 
equation \ref{eq:divb} can be derived from equations \ref{eq:ind1} and \ref{eq:ind2} 
by simple algebra.  There are also several integrals that can be derived from our
equations.  Equation \ref {eq:ind2} is already in the form of an integral.
Combined with equations \ref{eq:divb} and \ref{eq:continuity}, we easily find the 
following two integrals, in simplified form:
\bea
\bth &=& q \Wth, \label{eq:Bth_int} \\
\br &=& q \left(\Wr+\frac{2}{3} \right), \label{eq:Br_int} 
\eea

where $q$ is a constant.
A third integral can be obtained by inserting the first two into equations
\ref{eq:Psi} to \ref{eq:mom3} and seeking a relation between $\bbphi$ and $\Wphi$:
\be
\bbphi = q \Wphi + \Omega\sin\theta, \label{eq:Bphi_int}
\ee
where $\Omega$ is another constant.
Note that these integrals allow us to remove the self-similar magnetic field 
${\bf b}$ entirely from the equations to be solved.

Although it is not essential to the arguments of the present paper we
might reflect a little on the general significance of these
integrals. They are likely to be more general in fact than our
particular solution. In their present physical form they read:
\bea
 B_r &=&\frac{q}{\sqrt{G} t}\left(\frac{2}{3}\frac{r}{t}-v_r\right),\nn\\
 B_\theta &=&-\frac{q}{\sqrt{G}t}v_\theta,\nn\\
 B_\phi &=& \frac{\Omega r\sin\theta}{\sqrt{G}t^2}
 -\frac{q}{\sqrt{G}t}v_\phi.\nn
\eea
 If we rearrange these equations into vector form as 
\bea
{\bf v}-\frac{2}{3}\frac{{\bf r}}{t}&=&\frac{-\sqrt{G}t}{q}{\bf
  B}+\frac{{\bf \Omega}}{qt}\times {\bf r},\nn\\
& &\nn
\eea
where ${\bf\Omega}$ is along the axis of symmetry, then we can
recognize a kind of Ferraro's theorem (in the form $ {\bf v}
=const.\times {\bf B}+{\bf\omega}\times {\bf r}$ where ${\bf\omega}$
is the angular velocity of the field line) with time dependence. The
constants are time-dependent here and the material velocity is relative to a
freely-falling, zero-energy, Keplerian observer. One might expect the
time dependence to be different for different self-similar symmetries.

% ==================================================================================================

\section{Analytic Solution and Analysis}
\label{sec:Analysis}
\subsection{Boundary Conditions and Method of Solution}
\label{sec:method}

We now turn to the task of actually solving the self-similar equations presented in the
previous section, subject to all appropriate boundary conditions.  
With the help of the integrals given in equations \ref{eq:Bth_int} to \ref{eq:Bphi_int},
equations \ref{eq:continuity} to \ref{eq:Psi2} can be reduced to a system of 
equations involving only the self-similar velocity components ${\bf W}$, pressure $p$,
and their derivatives with respect to $\theta$.  Solving for the derivatives, 
the equations can be written as a dynamical
system in standard form, which specifies a boundary value problem in four variables on the 
interval $\theta \epsilon [0,\pi/2]$.

The boundary conditions at the symmetry axis and equatorial plane are as follows.  
The self-similar velocity components $\Wth$ and $\Wphi$ must vanish 
at $\theta=0$, where we also require that $\Wr>0$ for an outflow solution.  
The velocity component $\Wth$ must vanish at the equatorial 
plane $\theta=\pi/2$ for quadrupolar symmetry.  The ``dynamical'' 
component of the pressure $\p$ must be negative everywhere, since the pressure 
would increase radially outward otherwise, 
according to the self-similar form for the pressure given by equation \ref{eq:pform2}.
The total pressure can be made positive throughout the region of interest,
which is  bounded on the inside by the forming hydrostatic core and on the outside by the external
molecular gas that provides the source of the background pressure $p_0(t)$ in our model.
Note that we have demanded strict analyticity everywhere on the angular domain for the solutions
presented in this paper.  This restriction could, in principle, be relaxed by allowing a singular radial
velocity along the axis, in the spirit of FH1 and LHF.

The following symmetry conditions allow us to extend our solutions from $[0,\pi/2]$ to the 
full sphere $[0,\pi]$.  Under reflection about the midplane
($\theta\rightarrow \pi-\theta$), we assume that $W_\theta\rightarrow -W_\theta$, 
$b_r \rightarrow -b_r$, and $b_\phi \rightarrow -b_\phi$.  All other quantities remain unchanged upon reflection.
Note that the reversal of the magnetic field at the midplane requires the existence of a
current sheet in the equatorial plane, which is in accordance with the scenario discussed in Section \ref{sec:discussion}
for the generation of the quadrupolar field.  Note that the symmetry assumed here differs from
 our previous work, where we assumed that $W_\theta \rightarrow -W_\theta$ and $b_\theta \rightarrow -b_\theta$,
with all other quantities unchanged upon reflection.  In this case, continuity requires that both $W_\theta$ and $b_\theta$
must vanish on the midplane, thus requiring quadrupolar symmetry for all solutions.  
This is {\em not} required in general for the present model, which
admits solutions with dipolar symmetry in principle. Different equatorial boundary conditions are
required, notably $b_\theta\ne 0$ and continuous, but this is
permitted in principle by the time dependence.  We will, however, concentrate on the quadrupolar class of solutions in this paper.

We proceeded first with a numerical survey of the solutions in an
extended region of parameter space.  However, it gradually became clear from the extremely simple appearance of the 
solutions and one extremely robust result (see equation \ref{eq:two} below), that
an underlying analytical solution was at play.  Once we realized this, it was relatively
easy to discover the form of the analytical solution.  Given the parameters
$\mu$, $q$, and $\Omega$, we assume a solution of the following form and 
solve for the constants $c_1$ through $c_6$ in the resulting algebraic set of equations,
which are derived from the reduced system of equations involving only 
the self-similar velocity components and pressure:
\bea
\Wr(\theta) &=& c_1+c_2 \cos^2\theta \label{eq:Wr_form} \\
\Wth(\theta) &=& c_3 \sin(2 \theta) \label{eq:Wth_form} \\
\Wphi(\theta) &=& c_4 \sin\theta \label{eq:Wphi_form} \\
\p(\theta) &=& c_5 + c_6 \sin^2\theta. \label{eq:p_form} 
\label{eq:sincos}
\eea
After straightforward but tedious algebraic manipulations, it turns out that the constants $c_2$ through $c_6$
can all be expressed in terms of $c_1$ and the other three parameters mentioned above:
\bea
c_2 &=& -2-3 c_1 \label{eq:c1} \\
c_3 &=& -\frac{c_2}{2} \label{eq:c2} \\
c_4 &=& -\frac{c_2 q \Omega}{6\pi\mu (1+2 c_1)+q^2 c_2} \label{eq:c4} \\
c_5 &=& -\frac{2}{3}\pi\mu^2-\mu(1+2 c_1)(1+c_1) \label{eq:c5} \\
c_6 &=& \frac{1}{4\pi}\left\{-( c_4 q+\Omega)^2+2\pi\mu \left[ c_4^2-c_2 (1+c_1) \right] \right\} \label{eq:c6}
\eea
The parameter space is four-dimensional, with solutions to our boundary value problem 
completely specified by choosing $\mu$, $q$, $\Omega$,
and $c_1$.  

This solution clearly possesses quadrupolar symmetry, as defined above.  
We have checked this solution numerically for many 
values of the parameters.  In addition, our numerical analysis did not find any other classes of solution.  Thus, it appears 
that the analytic solution given above is the unique solution to our boundary value problem.
We suspect that other classes of solutions, including singular solutions and solutions with dipolar 
symmetry,  may exist if different boundary conditions are allowed. 

Not all of the solutions given by equations \ref{eq:Wr_form} to \ref{eq:c6} correspond to protostellar collapse 
and outflow solutions.
>From equation \ref{eq:Wr_form}, we note that outflow solutions require $c_1<0$ and $c_2>|c_1|$, so that 
material falls in near the equatorial plane and is ejected near the symmetry axis.  With the help of equation \ref{eq:c1},
this restriction implies that $c_1<-1$.  The opposite conditions provide us with
solutions in which the flow is reversed, with infall near the polar axis and expulsion 
near the equatorial plane.  The maximum value of the self-similar dynamical pressure is $c_5+c_6$.  
Thus, we require $c_5+c_6<0$, which keeps both the dynamical pressure and the outward 
pressure gradient negative, as discussed in Section \ref{sec:equations}.  There are, in fact, solutions with $p>0$ so that
the pressure increases with radius.  Such solutions might be physically interpreted as anisotropic collapse solutions
that are driven primarily by a sudden increase in the external pressure.  This sort of collapse might be reasonable in 
regions where star formation is initiated by a strong external shock.
However, we restrict ourselves to less exotic cases in this paper, where the pressure decreases radially outward.

\subsection{Characteristics of the Model and Exploration of the Parameter Space}
\label{sec:MC}
In this section, we discuss the general properties of the analytical solution given by equations \ref{eq:Wr_form} to \ref{eq:c6}.
We supplement this discussion by a rather complete exploration of the parameter space, in which we
allow the parameters $q$, $\Omega$, and $c_1$ to vary within pre-defined ranges given by
\bea
0 \le \mu \le 10^3    &  \qquad
-10^3 \le q \le 10^3  \nn\\
-10^3 \le \Omega \le 10^3 & \qquad
-10^3 \le c_1 \le -1.
\label{eq:edges}
\eea
We sampled these ranges normally in each parameter $p_i$ when $|p_i| \le 1$ and log-normally when $|p_i|>1$.
We find acceptable solutions right up to the edges of our parameter space given by
equation \ref{eq:edges}.  In principle, we could expand the region of 
our parameter space search even further than the ranges given above.
However, our choice to limit our parameter space is consistent with the spirit
of our self-similar analysis, where non-dimensional parameters are not usually expected to
be too many orders of magnitude greater or less than unity.

The \Alfven singularity in our self-similar equations occurs when $\ma=1$, where
\be
\ma^2\equiv\frac{4\pi\mu\Wth^2}{\bth^2}.   
\ee
Note that the usual \Alfvenic$~$ singularity is modified by our self-similarity ansatz  
so that only the $\theta$ components of the velocity and magnetic
field are involved (eg. Tsinganos \etal 1996, LHF).   
This can be simplified with the help of equation \ref{eq:Bth_int}:
\be
\ma^2=\frac{4\pi\mu}{q^2}.
\ee
The \Alfvenic mach number has a constant value for each solution, which does not
vary with either radius or angle.  Both super-\Alfvenic and sub-\Alfvenic models
are allowed, although there are no models where material passes through an \Alfven point.
Not, however, that  the special solution where $\ma=1$ everywhere is allowed.
Somewhat arbitrarily, we eliminate any solution for which $\ma<0.1$,
which corresponds to magnetic energy densities in $\bth$ greater than $50$ times the
kinetic energy density in $\Wth$.  This provides us with a convenient way to exclude
strongly sub-\Alfvenic models with unrealistically high magnetic field strengths.

There are several important conclusions that can be drawn directly from equations \ref{eq:Wr_form} to \ref{eq:c6}. 
>From equation \ref{eq:Wr_form}, and with the help of equation \ref{eq:c2}, 
the ratio of the outflow velocity to the infall velocity is given by
\be
\frac{\Wrout}{\Wrin}=-2-\frac{2}{c_1}.
\label{eq:max_velocity}
\ee
The ratio $\Wrout/|\Wrin|<2$ for all models, since $c_1$ is negative.
This upper limit on the net acceleration of the gas means that 
the axial outflow velocity is at most twice the equatorial inflow velocity
on a sphere of some given radius, at some instant of time.  
However, it is important to note that the velocity increases linearly with radius in our model, which is
very different from the model presented in FH1, where the velocity decreases with radius as $r^{-1/2}$.
Section \ref{sec:trajectories} shows that any given fluid particle escapes to radial infinity
along the outflow axis in the finite time required for the central singularity to develop (at $t=0$).
Thus, large outflow velocities are attained at some distance from the origin, near the time when the
singularity develops.  Realistically, the ejected material 
must encounter a shock with the external region at some finite radius, where the acceleration would presumably end.
Nevertheless, the possibility of a strong jet-like outflow remains, provided that the external medium is encountered at 
sufficient distance from the origin.

The early stage of outflow modelled in this paper is dominated by a quadrupolar circulation
pattern with an outflow along the symmetry axis.  It is very likely that the outflow would become even
more vigorous at a later stage of evolution, in which energy
is injected directly into the gas in the form of heat from the protostar.  
We have previously included such heating in a steady state
version of our quadrupolar flow model (FH1, LHF) and found that much 
higher velocities can indeed be achieved.  Thus, the present model might represent
the immediate predecessor in the evolutionary sequence leading up to our steady-state models.

\begin{figure}
\epsfig{file=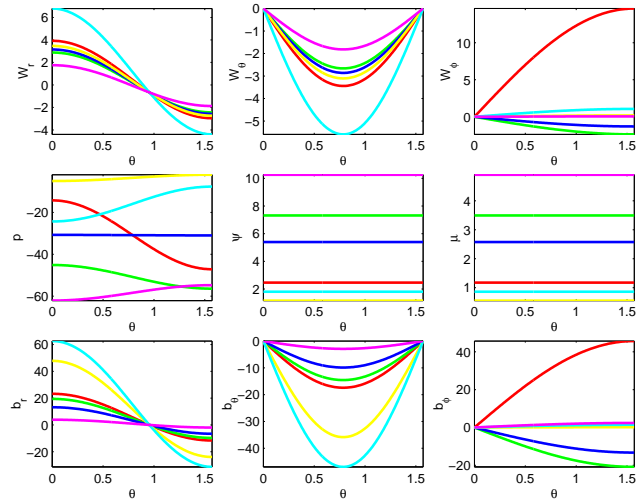,width=\linewidth}
\caption{We show an overlay of 6 typical solutions.  
%
% ****
Each model is colour coded for easy comparison between panels.
% ***Each model is coded with a combination of grayscale and line styles for easy comparison between panels.
%
}   
\label{fig:many}
\end{figure}

Figure \ref{fig:many} shows 6 of our solutions, chosen at random, overlaid on a single plot to illustrate
the large dynamic range of self-similar velocities, magnetic field strengths, and pressures allowed.  
The curves are 
%
% ****
colour-coded
% coded by grayscale and line style
%
for easy comparison between panels.
The large allowed range of initial conditions and protostellar properties
reflects the robustness of our model and of quadrupolar flow in general.

Now we examine the ratio of the radial magnetic field at the polar axis to the radial field at the
equator.  By combining equations \ref{eq:Br_int}, \ref{eq:Wr_form}, and \ref{eq:c2},
it is easy to show that 
\be
\frac{\Brout}{\Brin}=-2
\label{eq:two}
\ee
for {\em all} solutions.  This surprising result indicates that the flow is structured such that the radial component
of the magnetic field is always compressed to precisely the same extent.  It is interesting that we first noticed 
this result in outflow solutions obtained numerically using a shooting code.  This was one of the key results that 
suggested the existence of a general analytic solution to our boundary value problem, and led us to the solution presented in 
equations \ref{eq:continuity} to \ref{eq:Psi2}.

\begin{figure}
\epsfig{file=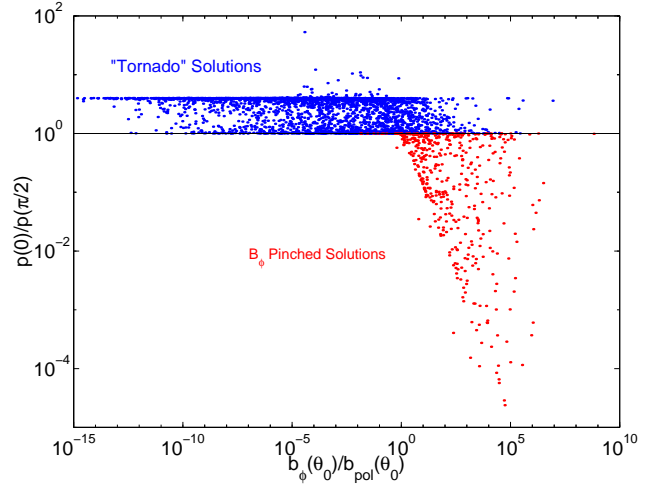,width=\linewidth}
\caption{Scatter plot of the ratio of the pressure at the axis over the pressure at the midplane, as a function of
$b_{\phi,max}^2/b_{r,max}^2$ for 5000 models.  $\p(0)/\p(\pi/2) >1$ corresponds to ``tornado'' type solutions,
while $p(0)/\p(0)/\p(\pi/2) <1$ models are pinched by the toroidal magnetic field. (Recall that $p<0$ - see Fig. 1). }
\label{fig:tornado}
\end{figure}

A careful examination of Figure \ref{fig:many} reveals that most of our models have a low pressure 
region near the symmetry axis, but some demonstrate the opposite behaviour.
This effect can be explained by consulting equations \ref{eq:p_form}, \ref{eq:Bphi_int}, \ref{eq:Wphi_form}, and \ref{eq:c6}.  Note that $c_6$ is the
self-similar pressure difference between the midplane and the polar axis:
\be
c_6=p(\pi/2)-p(0).
\ee
Note also that the term $(c_4 q+\Omega)$ in equation \ref{eq:c6} is the maximum value of the self-similar toroidal field $\bbphi$.  
Thus, a strong toroidal field tends to make $c_6$ negative, so that the pressure is highest at the polar axis.  Clearly, this 
compression is due to the radial pinch of the toroidal field toward the polar axis (e.g. Fiege \& Pudritz 2000a,b).
On the other hand, the constant $c_4$ in the second term is equal to the maximum rotational velocity, which acts in the opposite sense to 
push material away from the axis.  This centrifugal barrier produces a low pressure region near the axis when the toroidal field is
not sufficiently strong to counter its effect.  We refer to these models as ``tornado'' solutions, while we often 
refer to $\bbphi$ dominated
models as magnetically pinched.
A useful parameter to investigate these two types of behaviour is the 
ratio of the dynamical pressure at the the symmetry axis divided by the pressure at the midplane:
\be
\frac{\p(0)}{\p(\pi/2)}=\frac{c_5}{c_5+c_6}. \label{eq:pratio_def}
\ee
Since $p<0$, tornado-type solutions correspond to $\p (0)/\p (\pi/2) > 1$, while magnetically 
pinched solutions have the opposite behaviour, with $\p (0)/\p (\pi/2) < 1$.
In Figure \ref{fig:tornado}, we plot this quantity against the magnetic pressure ratio 
$\bphiturn^2/\bpolturn^2$, where $\bphiturn$ and $\bpolturn$ are respectively the self-similar toroidal 
field and poloidal field (defined by $b_{pol}=\sqrt{\br^2+\bth^2}$) components, evaluated at the angle
\be
\theta_0=\cos^{-1}\left[ \sqrt{-c_1/c_2} \right],
\ee
which divides the equatorial infall zone from the polar outflow zone (See equation \ref{eq:Wr_form}).
Since $c_1<-1$, equation \ref{eq:c1} implies that $\theta_0$ has a maximum possible value of $\cos^{-1}\sqrt{1/3} \approx 54.7^\circ$.
Note that the transition between tornado and magnetically pinched solutions occurs when $\bphimax^2/\brmax^2\approx 1$, 
as one would expect.  It is clear from the figure that most of the parameter space is filled with 
solutions of the tornado type.  Both super and sub-Alfv\'enic solutions are represented in this figure, with no apparent
discontinuity in behaviour.  Note the sharp boundaries that are visible in the distribution of tornado type solutions.
These are due to two limits that can be shown analytically from \ref{eq:pratio_def}, with the help of equations \ref{eq:c5}
and \ref{eq:c6}: $\p(0)/\p(\pi/2) \rightarrow 4$ as $c_1 \rightarrow -\infty$, and $\p(0)/\p(\pi/2) \rightarrow 1$ as $\mu\rightarrow \infty$.  

It is interesting to comment on the origin of the toroidal field component in our model.
The integral given in equation \ref{eq:Bphi_int} shows that the toroidal magnetic field
can be expressed as a the sum of two terms:
\be
\bbphi=\frac{\Wphi}{\Wr+2/3}\Br+\Omega\sin\theta
\ee
The first term is associated with toroidal field generated by $\Wphi$ twisting 
poloidal field lines,  while the second term can be identified with 
toroidal flux loops simply advected by the gas motions.

\subsection{Streamlines and Fieldlines}
\label{sec:lines}
\begin{figure}
\epsfig{file=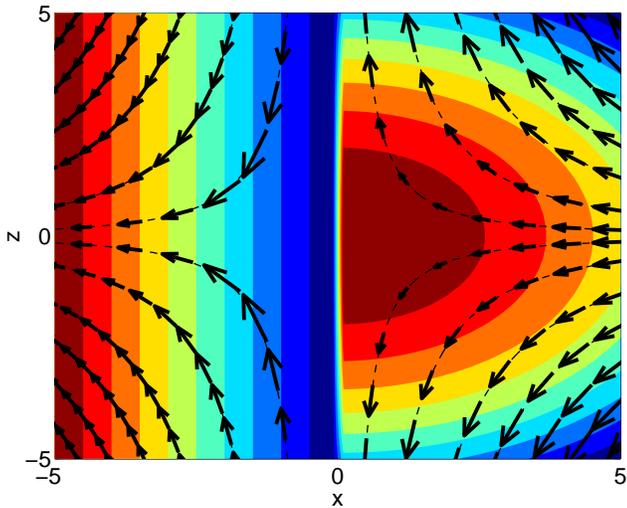,width=\linewidth}
\caption{We show poloidal field lines (left) and streamlines (right) for a typical solution.  These are overlaid   
on a 
colour 
% grayscale
%
plot of the toroidal field ``pressure'' $\bbphi^2/8\pi$ on the left, and the gas pressure on the right.  This
particular model is a ``tornado'' type solution.  The parameters of the model shown are as follows: $\mu=1.77$, $q=-6.11$, 
$\Omega=-14.56$, $c_1=-4.33$.}
\label{fig:streamfield}
\end{figure}

Figure \ref{fig:streamfield} is a split-frame figure showing the instantaneous stream (right hand side)
and field (left hand side) lines in the poloidal plane, overlaid with vectors whose lengths are
proportional to the magnitude of the poloidal velocity or magnetic field at each point. 
The stream and field lines are respectively given by solutions to the differential 
equations
\bea
\frac{dr}{r~d\theta} &=& \frac{\Wr}{\Wth} \label{eq:streamline} \\
\frac{dr}{r~d\theta} &=& \frac{\Br}{\Bth} \label{eq:fieldline}.
\eea
It is clear from the figure that the
stream and field lines in the poloidal plane are not parallel.  The stream and field lines are superimposed
over a 
%
% *****
colour
% grayscale 
%
representation of the pressure on the left, and the toroidal field strength on the right.

The path traced out by a fluid element in a time dependent model may not coincide with the instantaneous streamlines
because the streamlines themselves evolve in time.  However, an interesting property of our self-similar model is that 
a given fluid element followed in time actually obeys equation \ref{eq:streamline}, and hence moves
only along the instantaneous streamlines.  This behaviour arises
because time enters into each velocity component with the same power in equation \ref{eq:vform2},
and the components of $\W$ do not explicitly depend on time.  Thus the streamlines in the left panel of Figure
\ref{fig:streamfield} can also be interpreted as the paths of individual fluid elements.

\begin{figure}
\epsfig{file=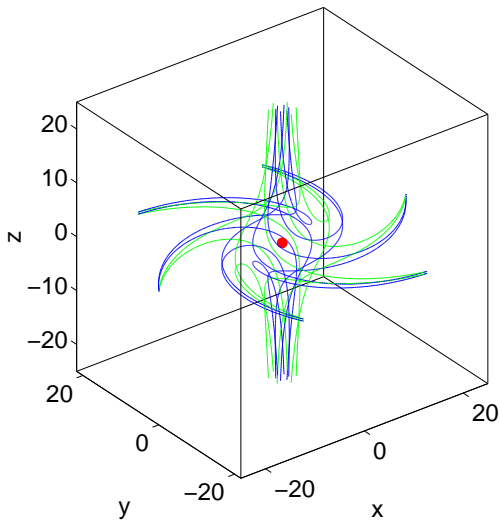,width=0.8\linewidth}
\caption{We show a three-dimensional view of several streamlines 
%
% ****
(blue) 
% (solid)
%
and field lines 
%
% ****
(green).
% (dashed).
%
Note that there is considerable rotation in flow, and that the streamline and field lines are not parallel,
indicating the presence of a Poynting flux.  The model shown is the same as the one in Figure \ref{fig:streamfield}.}
\label{fig:streamline3d}
\end{figure}

Figure \ref{fig:streamline3d} shows a three-dimensional representation of a several stream and field lines
equally spaced in angle and reflected about the midplane.  We have selected a model with a significant
amount of rotation for this figure, whose parameters given in the caption of figure \ref{fig:streamfield}).
The stream and field lines wraps around the origin as pressure gradients deflect 
the material out along the symmetry axis.  Note the similarity to the corresponding plot (Figure 5) in FH1.

\subsection{The Trajectories of Fluid Elements}
\label{sec:trajectories}
Some additional insight into the flow structure can be obtained by directly integrating
the trajectories followed by fluid elements.  It is trivial to integrate equation
\ref{eq:Wth_form}, with the help of equation \ref{eq:vform2}, to show that 
\be
\frac{\tan\theta}{\tan\theta_0}=\left( \frac{t}{t_0} \right)^{-2c_3},
\ee
where $\theta_0$ is the poloidal angle of a fluid particle at some initial time $t_0<0$.
Combining equations \ref{eq:c1} and \ref{eq:c2}, we find that
\be
c_3 =\frac{2+3 c_1}{2} < -1/2
\label{eq:c3b}
\ee
since $c_1<-1$ for all outflow solutions (Refer to the discussion following \ref{eq:c6}).
Thus, all fluid elements with $\theta_0<\pi/2$ go to the symmetry axis $\theta=0$ when the
central density becomes singular at $t=0$.  Fluid elements on the asymptotic streamline 
that starts at $\theta_0=\pi/2$ go to the origin at $t=0$.
Note that an initially spherical shell of fluid is transformed by the flow into an
thin ``needle-like'' distribution of material along the symmetry axis.

It is also possible to obtain an exact expression for the radial part of the trajectory by directly
integrating equation \ref{eq:streamline}, using the forms given in equations \ref{eq:Wr_form} and \ref{eq:Wth_form}.
We find that
\be
\frac{r}{r_0}=\left( \frac{\tan\theta}{\tan\theta_0}\right)^{c_1/(2c_3)} 
\left[
\left( \frac{\sin\theta}{\sin\theta_0}\right)
\left( \frac{\cos\theta_0+1}{\cos\theta+1} \right)
\right]^{c_2/(2c_3)}.
\ee
Therefore, $r \propto \theta^{(c_1+c_2)/(2c_3)}$ near the time $t=0$, when $\theta\rightarrow 0$.
The discussion following equation \ref{eq:c6} demonstrated that $c_1+c_2>0$ for all outflow 
solutions, and equation \ref{eq:c3b} indicates that $c_3<1$.  Thus, we find the intriguing result
that material escapes to infinity for all outflow solutions in the finite time required for the the central 
singularity to develop.  This confirms that our model can indeed generate strong, jet-like outflows, 
provided that they can escape to sufficiently large distances before encountering a shock with the 
external medium. 

\section{Discussion}
\label{sec:discussion}
\begin{figure}
\psfig{file=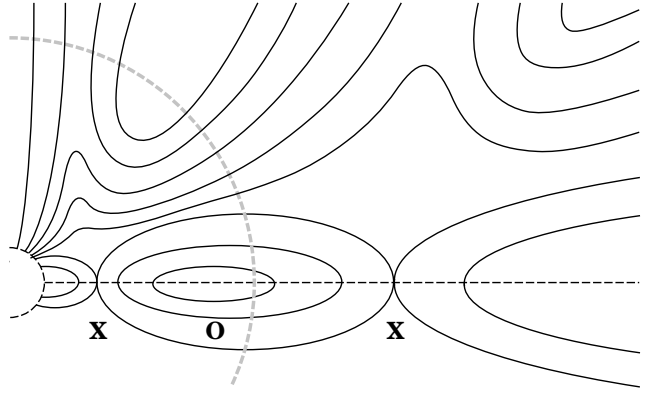,width=\linewidth}
\caption{We sketch the possible regions where quadrupolar stream and
  field lines may arise during the formation of a star. The solution
  presented in this paper is concerned with the inner region while
  that presented in our earlier papers concerns the outer region.}
\label{fig:cartoon}
\end{figure}

In this section, we deal mainly with the question of where the
preceding solution may be expected to apply.  The reader may find it helpful
to refer to Figure \ref{fig:cartoon} throughout this discussion, where we provide a sketch
of the magnetic field configuration that we describe below.

We proceed first with a qualitative discussion.  For simplicity, we
assume that the magnetic and angular momentum axes of a cold cloud
core coincide. The collapse of a molecular core may be due to
the inside-out collapse of a SIS (Shu 1977; GS), albeit one that
is threaded by a magnetic field.  Alternatively, it may be due to the 
loss of magnetic support due to ambipolar diffusion, 
or due to the damping of \Alfven waves or MHD turbulence.
The collapse might even result from the arrival of a sudden shock or 
torsional \Alfven wave (Hab\'e \etal 1991).
Whatever the cause of the collapse, it is likely that a nearly radial magnetic field
develops during collapse, as the field lines are dragged in with the collapsing gas.
These radial field lines will be oppositely directed across the equatorial plane.
It is in this plane (presumably near
the Keplerian point where the gas velocity may be relatively slow
compared to the magnetic diffusion speed) that one can expect the
central magnetic field to disconnect from the external field through 
the formation of an X type neutral point near the equator (e.g. Shu et al. 1994).  
The field lines should remain roughly radial outside of the equatorial X points,
especially near the midplane, as illustrated in Figure \ref{fig:cartoon}.

We must, however, also consider the pressure driven bounce of material
in opposite directions above and below the equator along the axis.
This rebounding material
forms a quadrupolar velocity field (FHI; Tomisaka,1998; LHF), 
which distorts the radial field locally into rising axi-symmetric poloidal arches
if the flow is super-Alfv\'enic.  As usual
there will also be a toroidal component of the magnetic field in these
arches.  These arches will also advect the toroidal component of the
field with the rising gas.  The key point is that these arches should develop first at small
radii, where we expect the most vigorous bounce.
Ultimately, we expect this flow pattern to lead to the formation of
quadrupolar velocity and magnetic fields consistent with the 
field structure that we have assumed in this series of papers.
However, a full quantitative analysis remains to be done.
The arches may themselves reconnect sporadically (through the formation of toroidal X lines), 
if the gas has finite resistivity so that ideal MHD does not strictly apply.
This would have the effect of producing magnetic toroids looped around the axis,
which would rise with the mean flow velocity.  After each such disruption the persistent
quadrupolar flow should reestablish the quadrupolar field.
These magnetic ejections may be associated with the
production of energetic particles and hence ``flaring''
(e.g. Montmerle \etal 2000 for the detection 
of significant X-ray emission from class I protostars), 
and are probably of interest in their own right
as features in the bipolar outflow.

The solution presented in this
paper applies at smaller radii than the radius of the
outer X point. 
As we have discussed above, the bounce at this outer point is created by a mixture
of rotational-magnetic support and radiative heating (e.g. LHF), whereas the bounce at 
the inner X point is caused by the forming hydrostatic core.
At this inner X point, we expect a
transition to a dipolar magnetic field (although a quadrupolar circulation
may still be present). Note that our equations may admit dipolar solutions, which could apply to the 
hydrostatic core at very small radii.  However, we have not yet found any such solutions.
Thus the settling solution that we have presented here 
is expected to lie between inner and outer X points.
Near the equator the field lines are actually wrapped around a magnetic O  
point. This equatorial XOX configuration may be a quite general
magnetic structure necessary to the process of star formation. It     
links the outer ``Keplerian'' disc to a more slowly growing inner hydrostatic
core. The core probably grows to become the star at the expense  
of the O-type envelope.

One can also get some feeling for the applicability of our solution by
looking at some numerical relations. We can rewrite equation
(\ref{eq:rhoform2}) to give $t=-\sqrt{(\mu/G\rho)}$ which shows that
$\sqrt{\mu}$ gives the time before the final singularity in units of
the (approximate) free-fall time. Let us suppose that the region in
which we are interested extends inward from about 1 au to the boundary
of the hydrostatic core. If this region has a mass $m$ in solar mass
units then $\rho\approx 2m\times 10^{-7}g~cm^{-3}$. Using $m=0.1
M_\odot$ for a low mass star, we arrive at $t\approx\sqrt{\mu}~years$. 
>From equations (\ref{eq:vform2}) and the solution (\ref{eq:Wr_form}),
we see that the inward equatorial velocity at 1 au is about $4
c_1/\sqrt{\mu}~km~s^{-1}$. Therefore if $c_1=-c\sqrt{\mu}$ where $c$
is of order one but is sufficiently large to make $p<0$ (see equation
\ref{eq:p_form}), then the bipolar velocity approaches $8c~km~s^{-1}$
and the equatorial velocity is comparable to the free fall velocity. 

It is also of interest to consider the magnitude of the magnetic field
in this illustration. At 1 au the equation (\ref{eq:bform2}) plus the
solution and our preceding assignments of the constants gives the
equatorial radial field to be of order $100~ gauss$. Of course this
ignores the total magnetic field which is a bit more difficult to
estimate because of the $\phi$ component and its relation to the
rotation of the system.   Note that these numbers are provided for 
illustrative purposes only.  Our models admits a large range of 
solutions that can describe a variety of protostellar objects.
   
\section{Conclusions and Summary}
\label{sec:summary}
We have developed a new model for the time dependent and anisotropic collapse
that occurs within the inner regions of a star-forming molecular cloud core.
By including the effects of self-gravity and MHD, our model provides a reasonably 
complete description of the dynamics on all scales between the inner hydrostatic core and 
an outer X point.  Our previous steady-state version of the model is expected to
apply external to the collapsing region modelled here, and possibly at later times.  
Remarkably, the collapse model presented in this paper admits an exact and
completely analytic solution.  We note that there are few other analytic solutions of this complexity in all
of MHD.  We summarize our solution here, for the reader's convenience:
\bea
v_r &=& -\frac{r}{t} (c_1+c_2 \cos^2\theta) \\
v_\theta &=& -\frac{r}{t} [c_3 \sin(2 \theta)]\\                
v_\phi &=& -\frac{r}{t} (c_4 \sin\theta) \\                
P &=& \frac{r_0^2}{G t^4}~p_0^\ast+\frac{r^2}{G t^4} [c_5 + c_6 \sin^2\theta] \\
\rho &=& \frac{1}{G t^2}~\mu,\\
B_r &=& \frac{r}{\sqrt{G}t^2} \left[q(c_1+c_2 \cos^2\theta+\frac{2}{3})\right],\\
B_\theta &=& \frac{r}{\sqrt{G}t^2} q c_3 \sin(2 \theta) \\
B_\phi &=& \frac{r}{\sqrt{G}t^2} (q c_4 + \Omega)\sin\theta \\
\Phi &=& \frac{2}{3}\pi \frac{r^2}{t^2} \mu,
\eea
where the four free parameters of the model are $\mu$, $q$, $\Omega$, and $c_1$.
The remaining constants in these expressions are given by 
equations \ref{eq:c1} to \ref{eq:c6}.  Note that the time $t$ starts large and
negative, and the model evolves until a central singularity forms at $t=0$.

The main point of this work is to demonstrate that infall and outflow can coexist and arise
naturally from our self-similar equations, with few additional assumptions.  The outflow that
arises is surprisingly vigorous, despite the lack of explicit internal heating in our model.
We find that the outflow velocity increases linearly with radius, and that
material in the outflow escapes to radial infinity in the finite time required for
the central singularity to develop.  More realistically, we expect the outflow to interact with the surrounding 
gas, outside of the self-similar region.  This would limit the velocity of the outflow, but high velocities
could still be obtained if the shock occurs much further out than the radius from which the 
material originated.  

Our model applies only at intermediate scales between two X type magnetic neutral points.
We expect a transition to a dipolar field internal to the inner X point, where the growing
hydrostatic protostellar core resides.  This region could, in principle, be modelled using the 
same set of equations we have used in the protostellar collapse region modelled here,
with boundary conditions consistent with a dipolar field.  The region exterior to the outer X point 
is probably characterized by longer dynamical timescales than the protostellar collapse region.
Once the overall quadrupolar flow structure is established by the collapse, the outer region might
be most appropriately modelled by the steady-state (or perhaps quasi-steady) version of this
model, which we have previously discussed in depth (FH1, LHF).  Future work on this class of models
may proceed in two alternate directions.  We may try to join together the three regions discussed
above, in piecewise fashion, to provide a more complete description of the simultaneous 
infall and outflow processes.  
Alternatively, we might turn to the more ambitious problem of solving the most general self-similar 
partial differential equations given in Section \ref{sec:equations}
(equations \ref{poiss} to \ref{phmom}) over the entire range of scales.

\section{Acknowledgements}
\label{sec:acknowledge}
This project was supported by a CITA/NSERC post-doctoral fellowship (JDF), 
an operating grant from NSERC (RNH), and by the combined support of NSF Grant
AST-0978765 and the University of Rochester's Laboratory for
Laser Energetics (TL).  A preliminary account of this work was
presented at the Cracow meeting on "Plasma Turbulence and Energetic
Particles in Astrophysics"(1999, M. Ostrowski and R. Schlickeiser,
eds.).  In addition, the authors would like to thank an anonymous referee for an
exceptionally thorough review, which helped us to strengthen our paper substantially.

\label{lastpage}

\end{document}